\documentclass{ws-ijmpe}
\usepackage[super,compress]{cite}
\begin{document}

\markboth{Jajati K. Nayak and Rupa Chatterjee}{Mean free path of photons in relativistic heavy ion collisions}

\catchline{}{}{}{}{}

\title{Mean free path of photons in relativistic heavy ion collisions}

\author{Jajati K. Nayak\footnote{jajati-quark@vecc.gov.in}}

\address{Variable Energy Cyclotron Centre \\
1/AF, Bidhan Nagar, Kolkata-700064, India }

\author{Rupa Chatterjee}

\address{Variable Energy Cyclotron Centre \\
1/AF, Bidhan Nagar, Kolkata-700064, India }

\maketitle

\begin{history}
\received{Day Month Year}
\revised{Day Month Year}
\end{history}

\begin{abstract}
Electromagnetic probes, such as photons and dileptons, play a key role in diagnosing the initial temperature of the hot and dense quark-gluon plasma (QGP) matter created in relativistic nuclear collisions at very high energies. This is due to their large mean free path $\lambda$, which allows them to escape the medium without significant interactions. Unlike hadronic particles, which experience multiple scatterings and are affected by the evolving medium, electromagnetic probes carry undistorted  information from the initial stages of the expanding system. In this work an attempt has been made to revisit the estimation of mean free paths of photons in QGP phase for a temperature range predicted by hydrodynamics for heavy ion collisions at $\sqrt{s_{NN}}=200$ GeV at RHIC and   $\sqrt{s_{NN}}=2.76$ TeV at the LHC. The mean free paths have been estimated for a plasma expanding via (1+1)D and (2+1)D hydrodynamical expansions. For the  (1+1)D case, photons with low energy ($E_{\gamma}<  0.2$ GeV) coming from a high temperature ($>250$ MeV) source are found to have shorter mean free path compared to the expansion scale of the system; while the high energy photons have always larger mean free paths. A similar qualitative nature of the mean free path has also been observed for a more realistic (2+1)D hydrodynamic model calculations although the $\lambda$ values are found to be larger on a quantitative scale compared to the (1+1)D case. 
\end{abstract}


\ccode{PACS numbers:}


\section{Introduction}
The mean free path of photons in relativistic heavy ion collisions is expected to be large compared to the size of the produced medium, enabling them to escape the hot and dense matter with minimal interaction and thus serve as a valuable probe of the initial temperature and other thermodynamic properties~\cite{phot1, phot2, phot3, phot4}. The matter produced at RHIC and LHC encounter both quark gluon plasma  and hadronic phases. It is thus important to estimate the mean free paths of photons in both QGP and hadronic media.

It is widely known that in such dense medium, photons have a high mean free path due to their weak interactions. They do not to undergo multiple scatterings, which allow them to carry information directly from their point of origin. The mean free path depends on factors like the photon’s energy, the temperature, and the density of the medium. High energy photons tend to have longer mean free paths due to decreased scattering probabilities, making them even more direct probes of early-stage collision dynamics.

Kapusta {\it et. al.}~\cite{kapustaprd} in 1991 have estimated the mean free paths of photons at various energies from both quark gluon plasma and hadron gas at different temperatures and found to be large. Based on the calculation they proposed that photons having energy of about one-half to several GeV can be a good signal of the formation of a quark-gluon plasma in high-energy collisions. That estimation gave an quantitative idea of photon mean free paths which justified the use of photon and lepton pair measurements as good probe of heavy ion collision to infer the initial temperature of the  produced system. The initial temperature in other words can justify or disprove the formation of quark gluon plasma. During that time heavy ion experiments were being carried out at CERN's Super Proton Synchrotron (SPS) and BNL's Alternating Gradient Synchrotron(AGS) facilities.  Later, heavy ion experiments at RHIC and at the  LHC provided significant evidences of the formation of quark gluon plasma in those collisions.

In the mean time significant advancement has been made  in hydrodynamic model calculations which provides a reasonably well  explanation of the bulk properties of the produced matter. No other calculations of the photon mean free path have been made since that time. The calculation of rate of the photon production which is an input to the estimation of mean free path has been modified significantly in last couple of decades. 
The state-of-the-art complete leading order~\cite{arnold2001} as well as NLO rates 
of thermal photons production from QGP~\cite{nlo} are available for quite some time now. There has been advancement in the photon production from the hadronic matter~\cite{trg} as well  which  includes the meson-meson and meson-baryon bremsstrahlung [see Ref.~\cite{phot2} and references therein for detail]. It is to be noted that Kapusta {\it et. al.}~\cite{kapustaprd} calculated the mean free path of photons ignoring the expansion of the matter. 

Here in this work an attempt has been made to revisit the calculation with expansion of the system and considering the rate of photon production upto leading order in $\alpha_s$. The  formalism for the calculation of mean free path has been discussed in the next section. Then in Sec. \ref{sec:rateandhydro} rate of photon production and hydrodynamics model calculations have been discussed. Then results are shown in Sec.~\ref{sec:results} and Sec.~\ref{sec:conclusions}  summaries the work.  
\section{ Mean free path of photon in heavy ion collisions \label{sec:meanfreepath}}
The relaxation time for a static system can be calculated from the rate equation as shown in \cite{kapustaprd}. The solution to the rate equation with initial zero photon density, at $t=0$, can be written as 
\begin{equation}
 \frac{dn}{d^3p}=\frac{dn^{eq}}{d^3p}\left(1-exp(-t/\tau) \right)
\end{equation}
The equilibration time is 
\begin{equation}
 \tau=\frac{dn^{eq}/{d^3p}}{dR/d^3p} 
\end{equation}
For an expanding QGP we can write 
\begin{eqnarray}
 \tau &=& \frac{dn/{d^3p}}{dR/d^3p} \nonumber\\
 &=& E\frac{dn}{d^3p}/E\frac{dR}{d^3p}
\end{eqnarray}
Where, $n$ is the number density of photons with momentum $p$ at any temperature $T$ and $EdR/d^3p$ is the rate of production and described in the next section. The mean free path $\lambda =c \times \tau$ and $c=1$ here.   

\section{Photon production and evolution of the system \label{sec:rateandhydro}}
In QGP, photon emissions are considered from Compton processes, $q(\bar{q}) g \rightarrow q(\bar{q}) \gamma$, annihilation processes $q \bar{q} \rightarrow g \gamma$ and bremsstrahlung processes $g q \rightarrow g q \gamma$, $q q \rightarrow q q \gamma$, $q q \bar{q}\rightarrow q q \gamma$ and $g q \bar{q} \rightarrow g \gamma$. The rate is given in \cite{arnold2001} 
\begin{equation}
 \frac{dN}{d^4x d^3p}=(\frac{1}{2\pi})^3 A(p)( ln[T/m_q(T) ] +\frac{1}{2} ln(2E/T)+C_{tot}(E/T))  
\end{equation}
Where, $E=p$ for mass less photons.

Thermal mass $m_q^2=4\pi \alpha_s T^2/3$,  $A(p)=2\alpha N_c\sum_s q_s^2\frac{m_q^2(T)}{E}f_D(E)$, $s$ is active quark flavors, $N_c$ =3, $q_s$ is the fractional quark charges and $f$ is the Fermi-Dirac distribution function. 
$C_{tot}(E/T)$ is given by the following expression, 
$$C_{tot}(E/T)=C_{2\rightarrow 2}(E/T)+C_{brem}(E/T)+C_{aws}(E/T) $$
Since all $C_i(E/T)$ contains non-trivial integrations, it is parametrised after numerical integration in \cite{renk2003} 
The parametrisation of rate upto order $\alpha_s$ \cite{renk2003} is used here. 
Au+Au collisions at 200A GeV and Pb+Pb collisions at 2.76A TeV are considered for both (1+1)D and (2+1)D ideal longitudinally boost invariant  hydrodynamical model evolution~\cite{jkn}. The initial parameters for the hydrodynamic framework are  tuned to reproduce the experimental data for charged particle multiplicity and  hadronic spectra both at RHIC and LHC~\cite{pramana}. A smooth initial density distribution has been considered for the initial state in central collisions and net baryon density is considered to be negligible at both energies. A lattice based EOS is taken for the (2+1)D model calculation with a transition temperature from QGP to hadronic phase of about 170 MeV~\cite{fcc}. 
\section{Results}\label{sec:results}

\begin{figure}
\begin{center}
\includegraphics[scale=0.4]{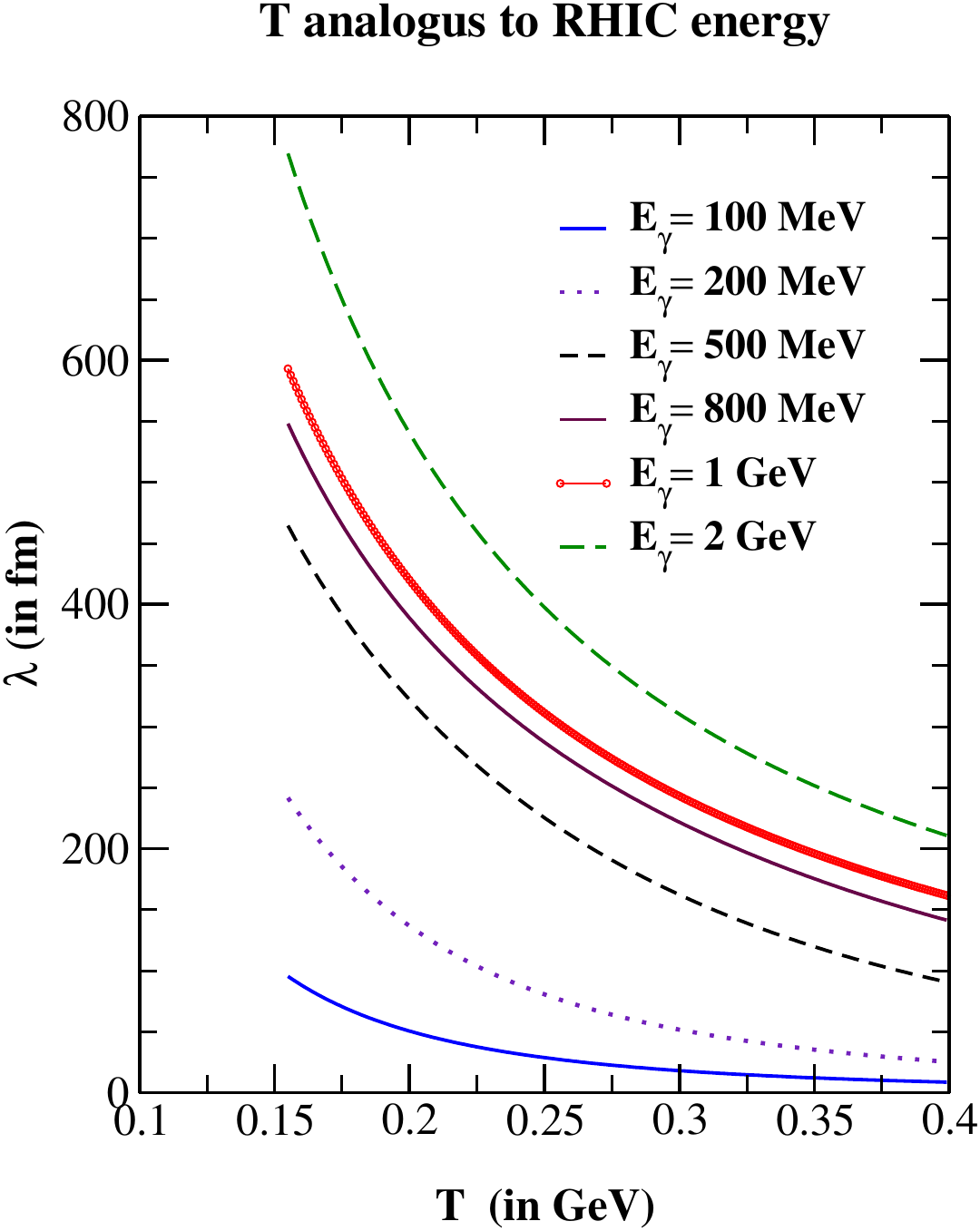}
\caption{ Mean free path of photons versus temperature coming with different energies, from (1+1) dimensional hydrodynamic calculation. }  
\label{fig1}
\end{center}
\end{figure} 
The photon mean path $\lambda$ from (1+1)D hydrodynamic calculation has been shown in Fig.~\ref{fig1}. The $\lambda$ values are plotted as a function of temperature T at different photon energies. The mean free path has been found to be extremely large around the QGP hadronic transition temperature and it decreases towards the hot and dense initial stage. A photon with larger energy results in  significant enhancement in the value of  $\lambda$ for a fixed temperature. 
The typical system size in Au+Au collisions is expected to be of the order of 10--20 fm which is way too smaller than the $\lambda$ of  the photons emitted with energy $\ge$ 0.2 GeV. 
\begin{figure}
\begin{center}
\includegraphics[scale=0.4]{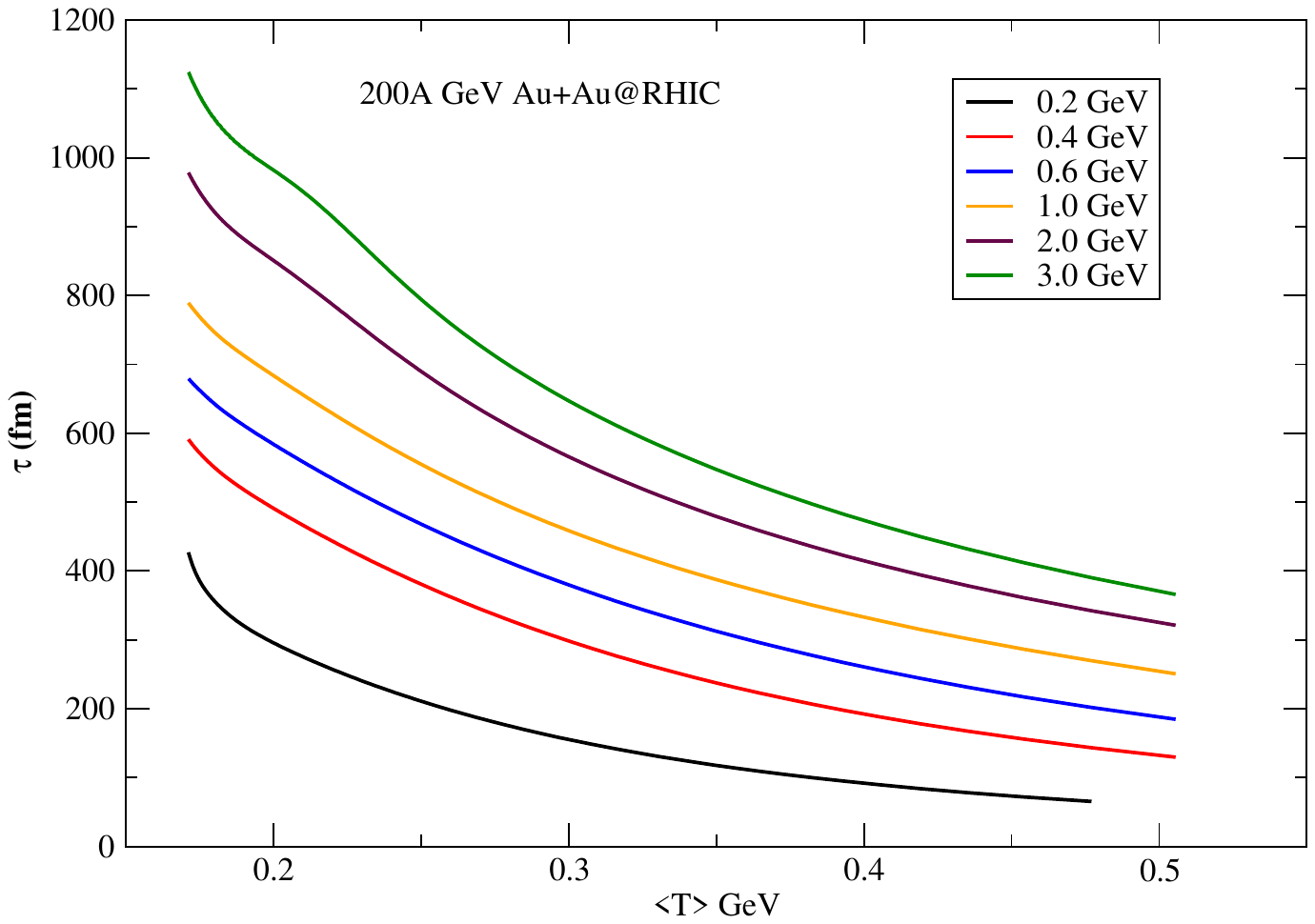}
\caption{ Mean free path of photons of different energies from expanding plasma from a (2+1)D hydrodynamical model calculation using rates from~\cite{arnold2001,trg}. }  
\label{fig2}
\end{center}
\end{figure} 

The results from a more realistic (2+1)D hydrodynamic model calculation at 200A GeV at RHIC has been shown in Fig.~\ref{fig2}. The temperature dependent mean free path values of the photons with energy rage 0.2 to 3 GeV in this case also shows similar qualitative nature as observed for the earlier case od (1+1)D calculation. However, the values of $\lambda$ are estimated to be larger (about 20\%) for the (2+1)D case. The generic nature of the $\lambda$ values are expected to be same even when initial state fluctuation and viscosity are included in the hydrodynamic model calculation. 

\section{Summary and Conclusions}
\label{sec:conclusions}
The photon mean free path $\lambda$ has been estimated using state of the art QGP photon rates and ideal hydrodynamical model framework in collisions of heavy ions at relativistic energies. 
These calculations reveal that $\lambda$ values are highly sensitive to both the temperature of the medium and the energy of the emitted photons. The results show that the photon mean free path is considerably larger than the size of the system, allowing them to escape with minimal interaction. These results show the importance of photons as reliable probes of the early thermal stages in heavy-ion collisions, as they provide a direct signal from the QGP with little distortion. However it is observed that photons from a high temperature phase coming with very low energy, of the order of a few hundred MeV, may have shorter mean free path compared to the expansion of the system.


\begin{thebibliography}{0}

\bibitem{phot1} E. L. Feinberg, \ Nuovo \ Cim. {\bf A34}, 391 (1976).
\bibitem{phot2} G. David, \ Rept. \ Prog. \ Phys. {\bf 83}, 046301 (2020).
\bibitem{phot3} J. Alam, B. Sinha, and S. Raha, \ Phys. \ Rept.  {\bf 273}, 243 (1996).
\bibitem{phot4}  P. Stankus, \ Ann. \ Rev. \ Nucl. \ Part. \ Sci. {\bf 55}, 517 (2005).
\bibitem{kapustaprd} J. Kapusta, P. Lichard and D. Seibert {\it Phys. Rev. D} {\bf 44} (1991) 2774.
\bibitem{arnold2001}P. Arnold, G.D. Moore, and L.G. Yaffe, {\it J. High Energy Phys.} {\bf 0112} (2001) 009.
\bibitem{nlo}J. Ghiglieri, J. Hong, A. Kurkela, E. Lu, G. D. Moore, D. Teaney,  JHEP  {\bf  05}, 010 (2013).
\bibitem{trg}S. Turbide, R. Rapp, and C. Gale, \ Phys. \ Rev. \ C   {\bf 69}, 014903 (2004).
\bibitem{renk2003}T. Renk {\it Phys. Rev. C} {\bf 67}(2003) 064901.
\bibitem{jkn} J. K. Nayak and R. Chatterjee [in preparation].
\bibitem{pramana}R. Chatterjee, Pramana  {\bf 95}, 15 (2021).
\bibitem{fcc}P. Dasgupta, S. De, R. Chatterjee, D. K. Srivastava,  \ Phys. \ Rev. \ C  {\bf 98}, 02491 (2018).
\end{thebibliography}
\end{document}